\documentclass[fleqn,usenatbib]{mnras}

\usepackage{newtxtext,newtxmath}
 
\usepackage[T1]{fontenc}

\DeclareRobustCommand{\VAN}[3]{#2}
\let\VANthebibliography\thebibliography
\def\thebibliography{\DeclareRobustCommand{\VAN}[3]{##3}\VANthebibliography}

\usepackage{graphicx}	
\usepackage{amsmath}	
\usepackage{amssymb}	

\usepackage{cleveref}

\defcitealias{Batten2021}{B21}

\title[FRBs as probes of feedback from AGN]{Fast radio bursts as probes of feedback from active galactic nuclei}

\author[A. J. Batten et al.]{
Adam J. Batten,$^{1,2}$\thanks{E-mail: abatten@swin.edu.au} Alan R. Duffy,$^{1,2,3}$
Chris Flynn,$^{1}$ 
Vivek Gupta,$^{1}$
Emma Ryan-Weber,$^{1,2}$
\newauthor
Nastasha Wijers$^{4}$
\\
$^{1}$Centre for Astrophysics and Supercomputing, Swinburne University of Technology, Melbourne, Victoria 3122, Australia \\
$^{2}$ARC Centre of Excellence for All Sky Astrophysics in 3 Dimensions (ASTRO 3D) \\
$^{3}$ARC Centre of Excellence for Dark Matter Particle Physics (CDM) Australia\\
$^{4}$Leiden Observatory, Leiden University, Niels Bohrweg 2, 2333 CA Leiden, The Netherlands\\
}

\date{Accepted XXX. Received YYY; in original form ZZZ}

\pubyear{2021}

\begin{document}
\label{firstpage}
\pagerange{\pageref{firstpage}--\pageref{lastpage}}
\maketitle

\begin{abstract}
Fast Radio Bursts (FRBs) are a promising tool for studying the low-density universe as their dispersion measures (DM) are extremely sensitive probes of electron column density. Active Galactic Nuclei (AGN) inject energy into the intergalactic medium, affecting the DM and its scatter. To determine the effectiveness of FRBs as a probe of AGN feedback, we analysed three different AGN models from the EAGLE simulation series. We measured the mean $\mathrm{DM}$-redshift relation, and the scatter around it, using $2.56\times10^8$ sightlines at 131 redshift ($z$) bins between $0 \leq z \leq 3$. While the $\mathrm{DM}$-redshift relation itself is highly robust against different AGN feedback models, significant differences are detected in the scatter around the mean: weaker feedback leads to more scatter.
We find $\sim 10^4$ localised FRBs are needed to discriminate between the scatter in standard feedback and stronger, more intermittent feedback models. The number of FRBs required is dependent on the redshift distribution of the detected population. A log-normal redshift distribution at $z=0.5$ requires approximately $50$\% fewer localised FRBs than a distribution centred at $z=1$. With the Square Kilometre Array expected to detect $>10^3$ FRBs per day, in the future, FRBs will be able to provide constraints on AGN feedback.
\end{abstract}




\section{Introduction}
Fast Radio Bursts (FRBs) are a newly discovered type of bright ($\sim$1 Jy), millisecond radio transient \citep{Lorimer2007, Thornton2013}. See \citet{Petroff2019, Petroff2021} and \citet{Cordes2019} and references therein for detailed reviews of FRBs.

In the short time since their discovery, FRBs already appear to be a promising tool for studying the cosmology and baryon structure of the Universe \citep{Bhandari2021}. Recently, using FRBs with localised host galaxies, \citet{Macquart2020} was successful in finally detecting the `missing baryons' in the low-redshift intergalactic medium (IGM). In addition, FRBs have also been proposed as probes for many areas of astrophysics including (but not limited to) constraining the timescales of helium reionisation \citep{Caleb2019,Linder2020,Bhattacharya2021}, measuring cosmological parameters \citep{Wu2020} and determining the strength of intergalactic magnetic fields \citep{Akahori2016,Hackstein2019}. In this Letter we propose using FRBs for studying the Active Galactic Nuclei (AGN) feedback efficiency of galaxies.

AGN feedback plays a pivotal role in the evolution of galaxies. AGN inject energy and material into the IGM due to the accretion of material onto their supermassive black holes \citep{Rees1984}. The energy released from this accretion is injected into the interstellar medium, heating the cold gas, and suppressing the next generation of star formation \citep{Scannapieco2005,Croton2006}. As a result, AGN have been shown to impact a large number of galaxy properties \citep{Schaller2015,Dubois2016,Pillepich2018}. Hence, constraining the strength and efficiency of AGN feedback is critical for galaxy formation models. A consequence of AGN feedback is the energy and material injected into the IGM changes the distribution of free electrons around galaxies. 

One of the key observable properties of FRBs is their dispersion measures (DMs) which are a measure of the integrated line of sight electron column density. The DMs of FRBs act as extremely sensitive probes of the low-density, ionised gas in the intergalactic and circumgalactic medium around galaxies and could potentially be used to constrain AGN feedback.

In \citet{Batten2021} (\citetalias{Batten2021} hereafter), we used the EAGLE simulations and measured the slope of the DM-$z$ relation (also known as the Macquart relation) and the width and shape of the scatter around the mean. In this Letter we expand on the work in \citetalias{Batten2021} and use the EAGLE simulations to determine whether FRBs are a suitable probe for measuring the efficiency of AGN feedback. In particular, the AGN contribution to the baryon cycle as it influences the amount of gas ejected over large length scales and recycling times.

This Letter is organised as follows. 
\Cref{sec:EAGLE} introduces the \mbox{EAGLE} simulations and the differences between the AGN feedback models. 
The production of dispersion measure maps is briefly described in \Cref{sec:Methods} (for full details see \citetalias{Batten2021}). 
In \Cref{sec:Results} we present the number of FRBs necessary to discriminate between the EAGLE AGN feedback models. We also determine redshifts of FRBs that are the most powerful for constraining AGN feedback in EAGLE. We summarise the conclusions in \Cref{sec:Discussion}.

\section{The simulations}\label{sec:EAGLE}
The Evolution and Assembly of GaLaxies and their Environments (EAGLE) simulations are suite of high-resolution, cosmological, $N$-body/hydrodynamic simulations (see \citealt{Schaye2015}, \citealt{Crain2015} and \citealt{McAlpine2016}). These simulations were run using a modified version of the smooth particle hydrodynamic (SPH) code \textsc{gadget-3}, last described in \citet{Springel2005}. EAGLE adopts a \citet{Planck2014} Lambda-Cold Dark Matter ($\Lambda$CDM) cosmology with the following parameters: $\Omega_m = 0.307$, $\Omega_\Lambda = 0.693$, $\Omega_b = 0.04825$, $H_0 = 67.77~\mathrm{km~s^{-1}~Mpc^{-1}}$, $\sigma_8 = 0.8288$, $n_s = 0.9611$ and $Y = 0.248$. The simulation cubes we discuss in this Letter were run with volumes of 50 cMpc per side and employed a maximum gravitational softening length of 0.7 pkpc. 
Process that occur on scales that are unresolved in EAGLE are implemented through a set of `sub-grid' models. 
These sub-grid models include: radiative gas cooling, hydrogen and helium reionisation, star formation, stellar evolution and mass-loss, star formation feedback, black holes and AGN feedback.
These sub-grid models are described in full in \citet{Schaye2015} and \citet{Crain2015}. 
However, since the AGN prescription is critical for this work, we will provide details of its implementation.

The AGN feedback in EAGLE is implemented through a single mode, where energy is stochastically injected to neighbouring particles as thermal energy (see \citealt{Booth2009}). Black hole particles of mass $10^5h^{-1}~\mathrm{M_\odot}$ are seeded in galaxy halos once they have grown to a mass of $10^{10}h^{-1}~\mathrm{M_\odot}$. The black hole particles are then allowed to grow by accreting neighbouring gas particles and through merging with other black holes. As the black hole accretes material it stores energy in a reservoir, $E_\mathrm{BH}$. After each timestep, $\Delta t$, $\Delta E = 0.015\dot{m}_\mathrm{accr}c^2 \Delta t$ is added to the reservoir where $\dot{m}_\mathrm{accr}$ is the gas accretion rate. Once the black hole has accreted enough material that the energy stored in its reservoir is enough to raise the temperature of at least one neighbouring particle by $\Delta T_\mathrm{AGN}$, each neighbouring particle is then stochastically heated. The probability of a particle being heated is
\begin{equation}
    P = \frac{E_\mathrm{BH}}{\Delta \epsilon_\mathrm{AGN} N_\mathrm{ngb} \left\langle m_g \right\rangle}\,,
\end{equation}
where $\Delta \epsilon_\mathrm{AGN}$ is the energy increment per unit mass corresponding to a temperature increment of $\Delta T_\mathrm{AGN}$, $N_\mathrm{ngb}$ is the number of neighbouring gas particles, and $\left\langle m_g \right\rangle$ is their mean mass. The black hole timestep was also limited to aim for probabilities of $P < 0.3$.

In this Letter, we have used three different AGN feedback models from EAGLE with identical initial conditions, box sizes and resolutions (L0050N0752). These include runs with feedback that is stronger but more intermittent (AGNdT9; $\Delta T_\mathrm{AGN} = 10^{9}$) relative to the Reference feedback (Ref; $\Delta T_\mathrm{AGN} = 10^{8.5}$), and a run without AGN feedback (NoAGN). Ref was calibrated to reproduce the redshift-galaxy stellar mass function, disk galaxy sizes and the black hole-stellar mass relation. Increasing $\Delta T_\mathrm{AGN}$ increases the energy required in the black hole reservoir before neighbouring particles are heated; meaning more energetic but more intermittent feedback \citep{Crain2015,Schaye2015}. We note that while NoAGN does not reproduce a realistic galaxy population, both Ref and AGNdT9 models do match observations. If FRBs are able to discriminate between Ref and AGNdT9, this would indicate they are able to provide a constraint not easily obtained from existing observations.

\section{Methods} \label{sec:Methods}
We have computed the DM-$z$ relation and produced DM PDFs at 131 redshift bins between $0 \leq z \leq 3.016$ for three simulations in EAGLE with different AGN feedback implementations: \mbox{RefL0050N0752}, \mbox{NoAGNL0050N0752} and \mbox{AGNdT9L0050N0752} (Ref, NoAGN and AGNdT9 hereafter). We have used the method as described in \citetalias{Batten2021} for producing the integrated DM maps (see also Section 2.2 of \citealt{Wijers2019}).
However the notable difference in this work is that we have used simulations with a smaller box size (50 cMpc per side instead of the 100 cMpc used in \citetalias{Batten2021}).
The smaller box size reduces the number of lines of sight to $16000^2$ ($2.56\times10^8$) keeping the area of each column fixed at $3.125^2$ ckpc$^2$.

We have summarised the method in \citetalias{Batten2021} as follows:
\begin{enumerate}
    \item We post-processed the EAGLE simulations to obtain number density of hydrogen (H) and helium (He) ions. 
    \item We produced maps of DM by summing along columns of fixed area to obtain the column density of H and He ions and converted them into electron column densities.
    \item To obtain continuous lines of sight out to redshift $z=3$, we produced 131 interpolated DM maps evenly spaced by 50 cMpc (the width of each box).
    \item Each line of sight was randomly assigned to new positions in the map to minimise periodic repeating structure before computing the cumulative sum along redshifts.
    \item Finally, we computed DM probability density functions for each of the 131 redshift bins.
\end{enumerate}

\section{Results}\label{sec:Results}

\subsection{The mean and scatter around the Macquart relation}
\begin{figure}
    \centering
    \includegraphics[width=\linewidth]{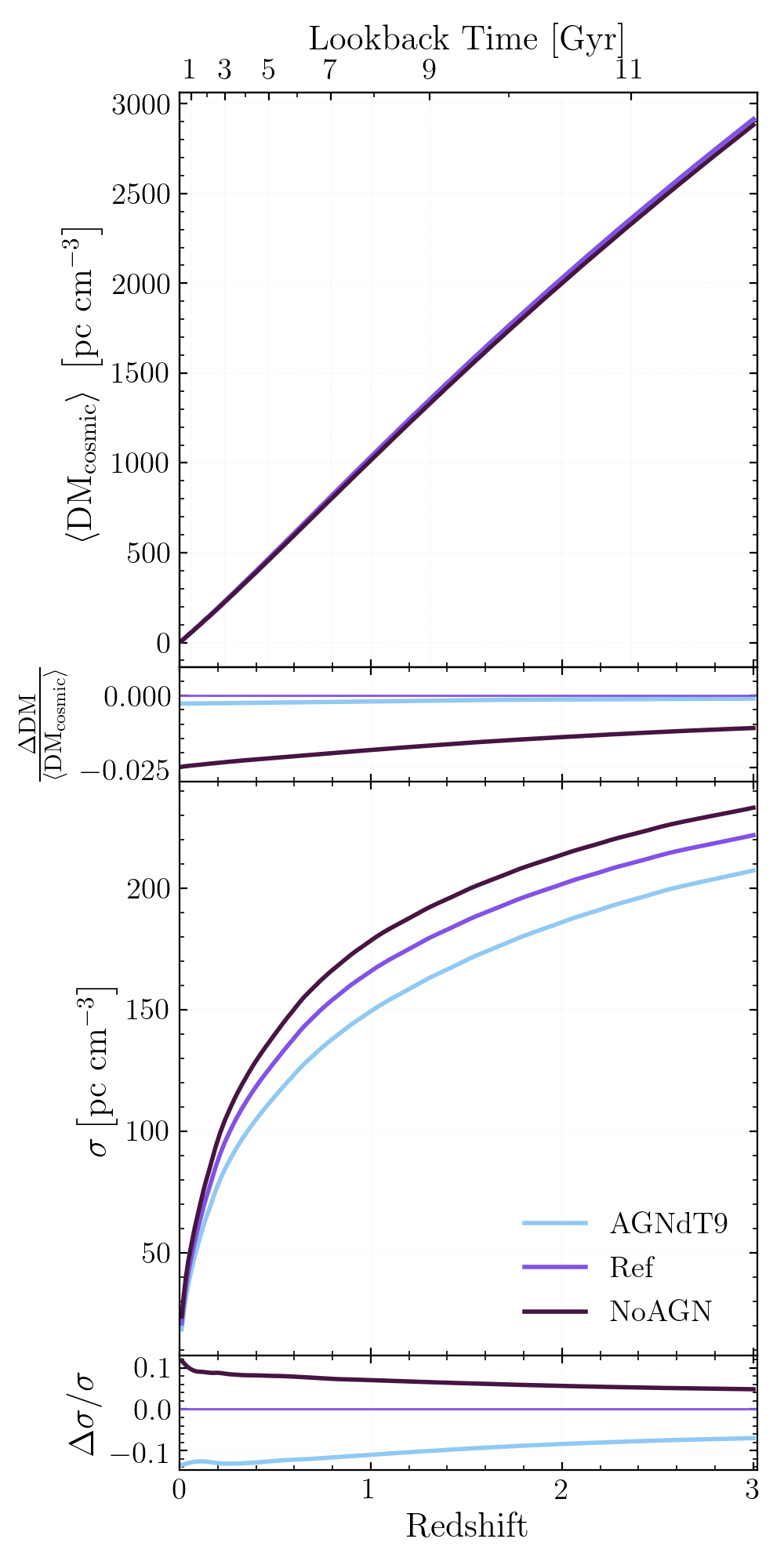}
    \caption{\emph{Top}: The mean $\mathrm{DM_{cosmic}}$ at each redshift for each for each of the simulations. Ref is the reference feedback model used in EAGLE, AGNdT9 is the stronger but more intermittent feedback model and NoAGN is the model with no AGN. \emph{Second}: The relative difference in the mean $\mathrm{DM_{cosmic}}$ to the Ref model. $\Delta \mathrm{DM} = \mathrm{DM_{cosmic,model}} - \mathrm{DM_{cosmic,Ref}}$. \emph{Third}: The standard deviation ($\sigma$) around $\left\langle \mathrm{DM_{cosmic}} \right\rangle$.  \emph{Bottom}: The difference in the standard deviation ($\Delta \sigma$) between simulations relative to Ref.}
    \label{fig:mean_sigma_v_redshift}
\end{figure}

In the first and second panels of \Cref{fig:mean_sigma_v_redshift}, we plot the mean cosmic dispersion measure ($\left\langle \mathrm{DM_{cosmic}} \right\rangle$) against redshift and the relative difference for the three AGN feedback models. We find that the DM-$z$ relations for the three AGN feedback models are practically indistinguishable, with relative differences (compared to Ref feedback) of $<0.5\%$ and $<2.5\%$ for AGNdT9 and NoAGN respectively. This indicates that the DM-$z$ relation is extremely robust against differences in feedback at all redshifts between $0 \leq z \leq 3$. This is result is also strongly converged with resolution and box size and is consistent with the DM-$z$ relations from \citet{Inoue2004}, \citet{Jaroszynski2019} and \citet{Macquart2020} (see \citetalias{Batten2021}).

In the third and fourth panels of \Cref{fig:mean_sigma_v_redshift}, we plot the standard deviation, $\sigma$, ($\sigma_\mathrm{var}$ in \citetalias{Batten2021}) around the mean $\left\langle \mathrm{DM_{cosmic}} \right\rangle$ and relative differences between models. We define $\sigma$ as
\begin{equation}
    \sigma^2(z) = \sum_{i=0}^{N_\mathrm{bins}} (\mathrm{DM}_i - \left\langle\mathrm{DM_{cosmic}}\right\rangle)^2 P(\mathrm{DM}_i|z)\Delta \mathrm{DM}_i \,,
    \label{eq:sigma_var}
\end{equation}
where $\mathrm{DM}_i$ is a bin value, $\left\langle\mathrm{DM_{cosmic}}\right\rangle$ is mean of the PDF at redshift $z$, $P(\mathrm{DM}_i | z)$ is the probability of a line of sight with $\mathrm{DM}_i$ at redshift $z$ and $\Delta \mathrm{DM}_i$ is the width of the bin. 

The NoAGN and AGNdT9 feedback models have the largest and smallest $\sigma$ at all redshifts between $0 \leq z \leq 3$. This indicates that weaker AGN feedback leads to a larger $\sigma$. This could be explained by the stronger feedback increasing the number of electrons in the IGM through baryons being ejected from galaxies more efficiently and the higher temperatures ionising the existing gas in the IGM. This change in number of electrons in the IGM is also not enough to significantly change the $\left\langle \mathrm{DM_{cosmic}} \right\rangle$ significantly.

The relative difference between the $\sigma$ in Ref and the other models is roughly flat, but slightly decreasing with increasing redshift. At low-redshift ($z < 0.5$), $\Delta\sigma/\sigma$ is approximately 10\%, but drops to 5\% by redshift $z=3.0$.

This variation in $\sigma$ between feedback models suggests that by measuring the scatter around the mean DM-$z$ relation of FRBs, it may be plausible to place observational constraints on AGN efficiency, particularly those used in simulations.   

\subsection{Required number of localised FRBs}
\begin{figure}
    \centering
    \includegraphics[width=\linewidth]{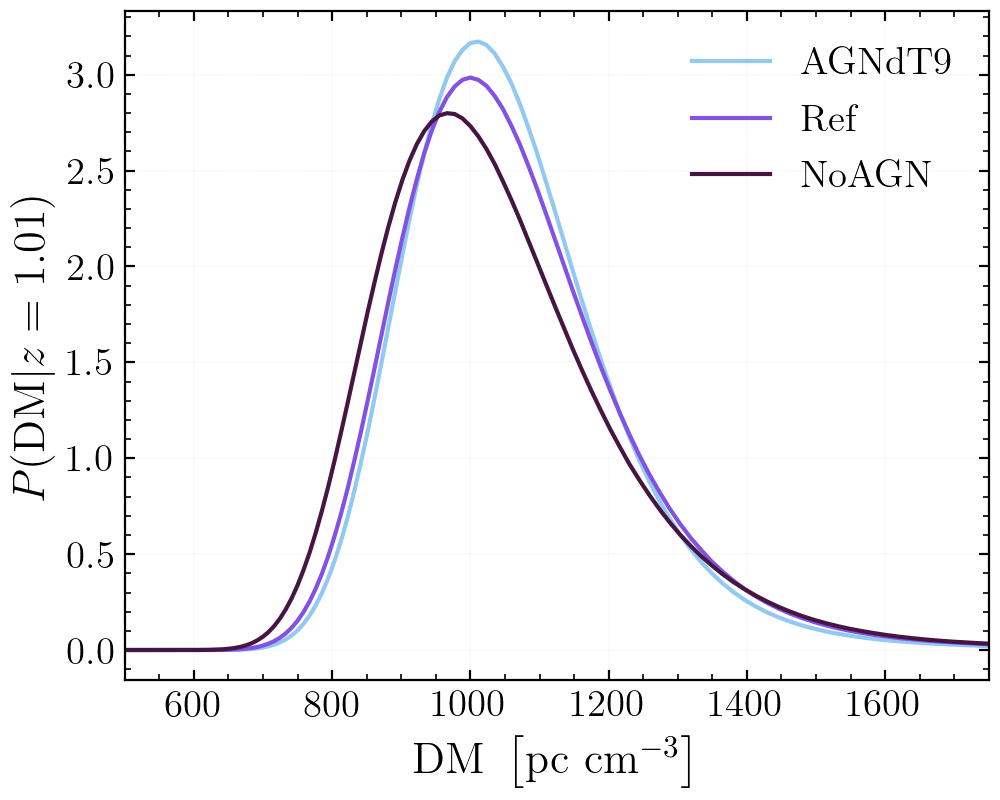}
    \caption{DM PDFs for each of the AGN feedback models at redshift $z = 1.01$. Ref is the reference feedback implemented in EAGLE, tuned to reproduce present day galaxies. AGNdT9 is AGN feedback that is more energetic and intermittent than Ref. NoAGN has no implementation of AGN feedback.}
    \label{fig:PDFs_redshift_1}
\end{figure}

To determine the number of localised FRBs that would be required to discriminate between these AGN models we use an Anderson-Darling k-sample test (A-D test). We have used an A-D test rather than a Kolmogorov–Smirnov test (K-S test) because the A-D test gives a higher weighting to the tails of the distribution. As shown in \Cref{fig:PDFs_redshift_1} and \citetalias{Batten2021}, the PDFs are significantly non-Gaussian and have long tails towards high DM values. When we refer to `localised' FRBs, we mean FRBs with an independently measured redshift.

We generated $10^6$ FRBs ($N_\mathrm{sample,total}$) for each of the AGN feedback models. The FRBs were drawn from a redshift distribution with their DMs sampled from the DM PDF of the feedback model. As the true underlying FRB redshift distribution is unknown, we have explored two models to compare the effects of different underlying populations.
The two redshift distributions we have explored are a log-normal distribution with a mean at $z=0.495$ ($\mu=5.987$, $\sigma=0.66$ in log-space), and a second log-normal distribution with a mean at $z=1$ ($\mu=6.690$, $\sigma=0.66$ in log-space). 
   
The log-normal distribution with a mean at $z=0.495$ is based on the best fit to the observed FRB DM distribution as measured by \citet{CHIME2021}. The observed DM distribution appears to fit a log-normal distribution with a mean $\left\langle \mathrm{DM} \right\rangle = 495~\mathrm{pc~cm^{-3}}$ and $\sigma=0.66$ (in log-space). We used the crude approximation that $\mathrm{DM} = 1000z$ to fix the mean of our redshift distribution at $z=0.495$, also using $\sigma=0.66$. We note that we have not corrected for the Milky Way and host galaxy components however those corrections are small (both the Milky Way and host galaxy components are approximately 30 $\mathrm{pc~cm^{-3}}$). If we were to remove the Milky Way and host galaxy components, this would be equivalent to using a log-normal distribution with $\mu=5.85$ and leads to a small change on the estimation of $N_\mathrm{loc}$ ($<30$\% reduction). 

We emphasise that this model is to simply broadly reproduce a realistic DM distribution. The second log-normal distribution is the same as the first but with the mean shifted to $z=1$.

We performed an A-D test on a sub-sample of $N_\mathrm{loc}$ FRBs from the larger $N_\mathrm{sample,total}$ sample. Here $N_\mathrm{loc}$ ranges from $10-10^5$. We have ensured that the FRBs selected in both models are from the same redshifts (although independently sampled from the DM PDFs). We repeat the A-D test $10^4$ times, including resampling $N_\mathrm{loc}$ from $N_\mathrm{sample,total}$ each time. Finally, we calculated the mean and standard deviation of A-D test statistic. We note that in rare cases when the A-D test returns a p-value of $p<0.001$, the test stat is capped at 6.546, leading to the mean test statistic being very slightly underestimated\footnote{This is due to the functionality in \href{https://docs.scipy.org/doc/scipy/reference/generated/scipy.stats.anderson_ksamp.html}{scipy.stats.anderson\_ksamp}.}. However, this makes no significant difference on our results.

\begin{figure}
    \centering
    \includegraphics[width=\linewidth]{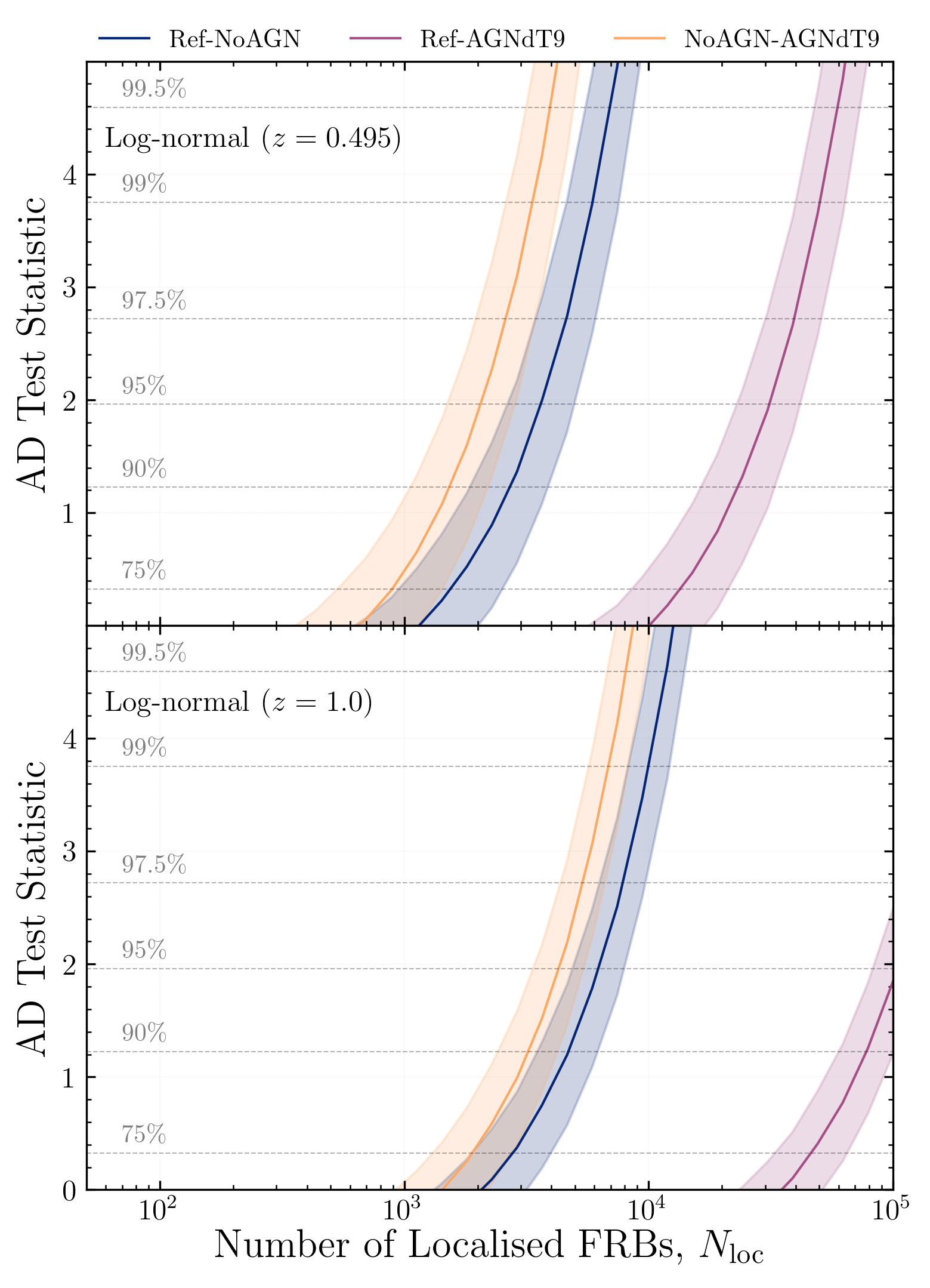}
    \caption{\emph{Top}: The number of localised FRBs required to discriminate between AGN feedback models assuming an underlying FRB population follows a log-normal distribution with a mean redshift of $z=0.495$ ($\mu=5.987, \sigma=0.66$ in log-space). The solid coloured lines are the mean significance level achieved using $10^4$ Anderson-Darling k-sample tests and the shaded regions are the $1 \sigma$ (68\%) confidence intervals. Ref is reference feedback, NoAGN is no AGN feedback, AGNdT9 is stronger but more intermittent AGN feedback. The grey horizontal dotted lines are the A-D test statistic converted into a significance level. \emph{Bottom}: The same as the middle panel except the mean of the log-normal distribution has been shifted to $z=1.0$ ($\mu=6.690, \sigma=0.66$ in log-space).}
    \label{fig:num_localised_frbs}
\end{figure}

In \Cref{fig:num_localised_frbs} we plot the A-D test statistic against $N_\mathrm{loc}$. These plots shows the number of localised FRBs that are required distinguish between the PDFs of the model pair; up to a given significance level. The top and bottom panels are the log-normal centred at $z=0.495$ and log-normal centred at $z=1$ redshift distributions respectively. The solid coloured line is the mean A-D test statistic after $10^4$ tests and the shaded regions are the 1$\sigma$ (68\%) confidence intervals. The grey horizontal, dotted lines are the significance levels of the A-D test statistic. For the purposes of this Letter, we define that it is possible to distinguish between the two if the A-D test statistic $> 1.961$ (95\% significance level).

\begin{figure}
    \centering
    \includegraphics[width=\linewidth]{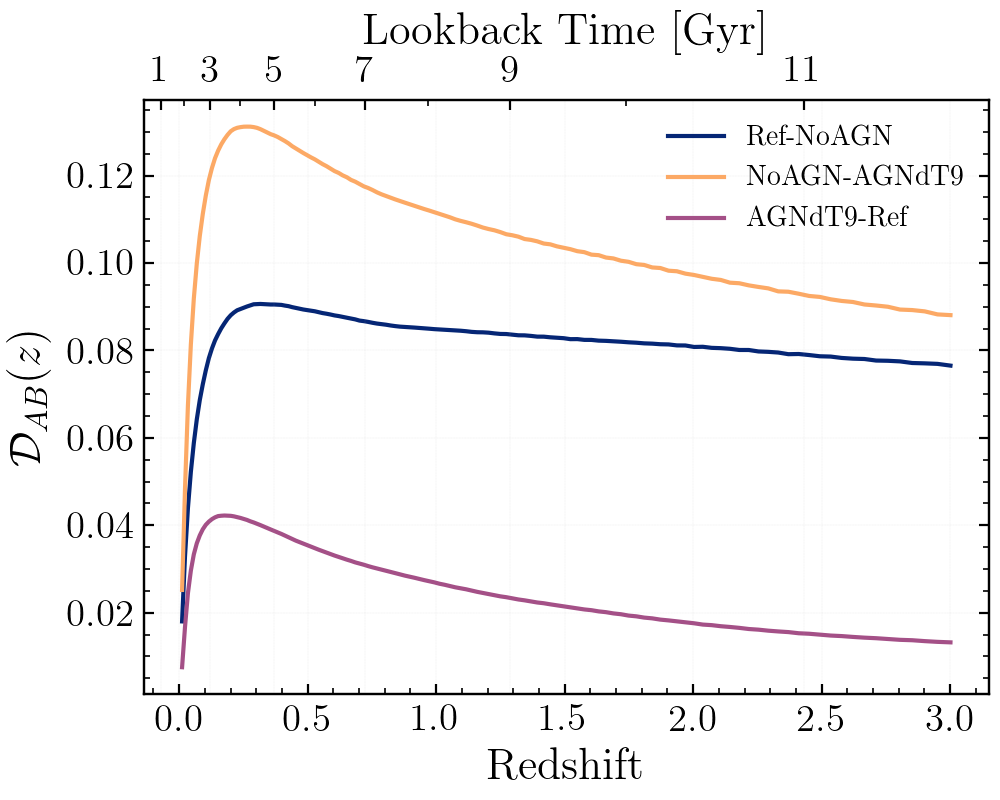}
    \caption{The KS distance metric, $\mathcal{D}_\mathrm{AB}$, between the PDFs of each model pair. Larger values of $\mathcal{D}_\mathrm{AB}$ indicate the models PDFs are more different. The peak value of $\mathcal{D}_\mathrm{AB}$ occurs at redshift $z= (0.32, 0.26,0.18)$ for the Ref-NoAGN, NoAGN-AGNdT9 and Ref-AGNdT9 model pairs respectively.}
    \label{fig:cdf_distance}
\end{figure}

The two feedback models that can be distinguished with the least number of FRBs are NoAGN and AGNdT9. This is an intuitive result because these models are the `most different' from each other (i.e No AGN vs strong AGN, see \Cref{fig:cdf_distance}) and thus distinguishing between them should should require the least FRBs. Of the order $10^{2.6} - 10^{3.6}$ (depending on the redshift populations of FRBs) FRBs are required to discriminate between NoAGN and AGNdT9. A similar but slightly higher number of FRBs is required for NoAGN and Ref.

However, discriminating between the two different strengths of AGN feedback requires significantly more FRBs. The log-normal ($z=0.495$) redshift distribution required approximately $10^4$ localised FRBs to discriminate between the Ref and AGNdT9 models. The number of FRBs increases by an order of magnitude to $10^5$ if the redshift distribution is shifted to $z=1.0$. At a rate of 5 FRB localisations and redshifts per day, it would only take approximately 5.5 years to reach $10^4$ FRBs. With the upcoming Square Kilometre Array (SKA) expecting to detect $10^3$ FRBs per day with redshifts $z > 1$ \citep{Hashimoto2020}, it will be likely that there will soon be enough FRBs to place constrains on AGN feedback.

\subsection{Optimal redshift for localised FRBs}
We now determine the redshift at which FRBs are most effective at constraining AGN feedback. A proxy for the redshift at which FRBs are the most effective, i.e. when the DM PDFs are most different, is a measure of their `distance', $\mathcal{D}_\mathrm{AB}$. We define the distance $\mathcal{D}_\mathrm{AB}$ between the PDF of model $A$ and model $B$ as,
\begin{equation}
    \mathcal{D}_\mathrm{AB}(z) = \mathrm{Max} \left(\left|\mathrm{CDF_A}(z) - \mathrm{CDF_B}(z)\right| \right)\,,
\end{equation}

where $\mathrm{CDF_A}(z)$ and $\mathrm{CDF_B}(z)$ are the DM CDFs for AGN feedback model $A$ and $B$ at redshift $z$ respectively. The $\mathcal{D}_\mathrm{AB}$ between two models is equivalent to the K-S statistic between the model PDFs. A larger value of $\mathcal{D}_\mathrm{AB}$ indicates that the models are more different. 

In \Cref{fig:cdf_distance}, we plot $\mathcal{D}_\mathrm{AB}$ for each pair of models at each redshift. For each model pair, $\mathcal{D}_\mathrm{AB}$ rapidly increases with distance, peaking at redshift $z= (0.32, 0.26,0.18)$ for Ref-NoAGN, NoAGN-AGNdT9 and Ref-AGNdT9 respectively, and then declining with a long tail. The peaks indicate that FRBs at redshift $z\approx0.2$ would be the most powerful in regards to constraining AGN feedback.

\section{Conclusions}\label{sec:Discussion}
In this Letter we have explored the possibility of using FRBs to probe and constrain the AGN feedback of galaxies. In particular, feedback that contributes to the baryon cycle; the amount of gas ejected and the timescales for removal and recycling of material. Our analysis of three different AGN feedback models in the EAGLE simulation suite has led to the following conclusions:
\begin{itemize}
    \item The DM-$z$ relation (Macquart relation) for FRBs is extremely robust against changes to AGN feedback models. The three different AGN feedback models (reference feedback, no feedback, stronger but more intermittent feedback) tested produced extremely similar DM-$z$ relations (<2.5\% difference). 
    \item The scatter around the DM-$z$ relation varies between feedback models. At low redshifts ($z<0.5$) the standard deviations differ by as much as 10\%, but the differences decreases with increasing redshift.
    \item Using an Anderson--Darling k-sample test we determined that approximately $10^4$ localised FRBs with redshifts will be needed to discriminate between the Reference feedback in EAGLE and the stronger \& intermittent feedback. At a rate of 5 localised FRBs per day, it will only take 5-6 years to reach $10^4$ FRBs.
    \item The differences between the DM PDFs peaks at redshift $z \approx 0.2$. This suggests that FRBs at redshift $z \approx 0.2$ provide the highest constraining power on AGN feedback models.
\end{itemize}
In the future we will need more large box cosmological simulations with varying feedback models to created `grids' of parameter space that can be constrained by FRBs.

\section*{Acknowledgements}
We acknowledge the Wurundjeri People, the traditional owners of the land upon which Swinburne University of Technology (Hawthorn) is located. This research was supported by the Australian Research Council Centre of Excellence for All Sky Astrophysics in 3-Dimensions (ASTRO 3D), through project number CE170100013. This work was performed on the OzSTAR national facility at Swinburne University of Technology. The OzSTAR program receives funding in part from the Astronomy National Collaborative Research Infrastructure Strategy (NCRIS) allocation provided by the Australian Government. This research has made use of NASA's Astrophysics Data System and software including: \texttt{matplotlib} \citep{Hunter:2007}, \texttt{SciPy} \citep{Virtanen_2020}, \texttt{IPython} \citep{PER-GRA:2007}, \texttt{Astropy} \citep{2013A&A...558A..33A, 2018AJ....156..123A}, \texttt{NumPy} \citep{van2011numpy, harris2020array}, \texttt{Pandas} \citep{McKinney_2010, McKinney_2011} and \texttt{CMasher} \citep{CMasher}.

\section*{Data Availability}
The data in the figures including plotting scripts and simulations will be shared on a reasonable request to the corresponding author. The particle data for the EAGLE simulations are available at \href{http://icc.dur.ac.uk/Eagle/database.php}{http://icc.dur.ac.uk/Eagle/database.php}



\bibliographystyle{mnras}
\bibliography{references} 








\bsp	
\label{lastpage}
\end{document}